\documentclass[12pt]{iopart}
\usepackage{graphicx}
\begin{document}

\title{A new possibility for light-quark Dark Matter}

\author{M. Bashkanov}
\address{ Department of Physics, University of York, Heslington, York, Y010 5DD, UK}
\ead{mikhail.bashkanov@york.ac.uk}

\author{D. P. Watts}
\address{ Department of Physics, University of York, Heslington, York, Y010 5DD, UK}
\ead{daniel.watts@york.ac.uk}

\vspace{10pt}
\begin{indented}
\item[]July 2019
\end{indented}

\begin{abstract}

Despite many decades of study the physical origin of  "dark matter" in the Universe remains elusive. In this letter we calculate the properties of a completely new dark matter candidate - Bose-Einstein condensates formed from a recently discovered bosonic particle in the light-quark sector, the $\mathbf{ d^*(2380)}$ hexaquark.  In this first study, we show stable $\mathbf{ d^*(2380)}$ Bose-Einstein condensates could form in the primordial early universe, with a production rate sufficiently large that they are a plausible  new candidate for dark matter. Some possible astronomical signatures of such dark matter are also presented. 
\end{abstract}

%
%
\submitto{\JPG}
%
%
%

\section*{Introduction}

The physical origin of dark matter (DM) in the universe is one of the key unsolved questions for physics and astronomy. There is strong indirect evidence for the existence of such matter\cite{DMrev} from measurements of cosmic primordial radiation, anomalies in the radial dependence of galactic rotational curves and gravitational lensing. Despite its apparently pivotal role in the universe the physical origin of DM remains unknown, with significant research focused on beyond standard model (yet currently undiscovered) particles such as axions, sterile neutrinos and weakly interacting massive particles (WIMPs)\cite{DMrev}. Recent advances in experimental searches have now eliminated significant fractions of the parameter space for WIMP candidates and the initial motivation as a full solution to the dark matter problem appears weaker. For example, the upper abundance limits derived from the latest direct detection WIMP searches\cite{DMXenon} appear incompatible with many of the supersymmetric WIMP DM models. 

 As well as the exotic searches described above the possibility of DM deriving from light-quark ingredients remains appealing as the mass of the "remaining" visible matter in the universe is predominantly in this form, albeit mainly as fermionic (half-integer spin) configurations of light u,d quarks such as the protons and neutrons within atomic nuclei. The possibility for additional bosonic (integer-spin) forms of light-quark matter have been investigated, particularly 6-quark (hexaquark) systems such as the double-strange quark containing "H-dibaryon"\cite{HDbar}. However, despite recent ongoing work\cite{HDM} such postulates went out of favour as a H-dibaryon with the desired properties for a significant contribution to DM was ruled out experimentally. 

In this letter we present a new possibility for light-quark based DM, motivated by the recent discovery of the $d^*(2380)$ hexaquark. The properties of the $d^*(2380)$ have been established in recent years following its first experimental  observation\cite{mb,MB1,MBC,TS1,TS2,MBA,MBE1,MBE2,BCS}. It has a mass of $M_{d^*}=2380$~MeV, vacuum width $\Gamma = 70$~MeV and quantum numbers $I(J^P)=0(3^+)$ and made of six light quarks; 3 u-quarks and 3 d-quarks. The spins of the quarks are aligned and QCD based approaches, such as Chiral quark models, predict a highly compact structure, smaller than a single proton\cite{DBT9}. It therefore offers a completely new {\em bosonic} and {\em isoscalar} configuration into which light-quark matter can form.

In this letter we present the first estimates of the stability, properties and primordial production rates of a $d^*(2380)$-BEC, employing a range of simple theoretical ansatz and physically motivated assumptions based on our current knowledge of the $d^*(2380)$ and its interaction. We discuss how charge-neutral bound systems based on a $d^*(2380)$-BEC core are a new DM candidate and  outline potential astronomical signatures that would arise from such systems. 

\section{\label{sec:dic}Estimates of $d^{*}(2380)$-BEC stability}
To estimate the stability of the $d^{*}(2380)$-condensate we take ansatz used previously in the evaluation of BEC properties and adapt these for the case of the $d^{*}(2380)$. The simplest such approach for BECs is a modified liquid drop model. For a standard isospin symmetric nucleus the liquid drop model gives the nuclear binding energy, {\it $B_{LD}$}, per nucleon as 
\begin{equation}
B_{LD}/A=a_v-a_s\cdot \frac{1}{A^{1/3}}-a_c\cdot \frac{Z(Z-1)}{A^{4/3}}    
\end{equation}
where {\it A }is the mass number, {\it Z} the proton number, $a_v$ the volume coefficient, $a_s$ the surface coefficient and $a_c$ the Coulomb coefficient.

To model a BEC the terms in the drop model need to be modified for {\it bosonic} rather than nucleonic constituents. We follow the ansatz used for the modelling of other bosonic condensate systems, such as linear chain $\alpha$ particle condensates\cite{alcon}. The surface term is assumed obsolete; in a BEC all constituents have a common wavefunction forming a single object with binding independent of radial position. This also influences the form of the volume term; the equivalent interactions between all BEC constituents give a proportionality to $A(A-1)$, in contrast to nuclear matter, where the dominance of nearest neighbour interactions produces a proportionality of binding with {\it A}. The lack of surface effects in a BEC also influences the form of the Coulomb term, as it is then energetically favourable for the condensates to form a linear chain (where the Coulomb repulsion is minimised), rather than a sphere as assumed in Eq.1. Based on these modifications the drop model then gives the binding energy per $d^*(2380)$ ($B/D$) in a linear (lin) and spherical (sph) BEC configurations as:


\begin{eqnarray}
B/D=a_v'\cdot(D-1)-a_c'\cdot \frac{(D-1)}{D};~~a_c'[MeV] = 0.064\cdot\frac{\rho}{\rho_0};~~ (lin)   \\ 
B/D=a_v'\cdot(D-1)-a_c'\cdot \frac{(D-1)}{D^{1/3}};~~a_c'[MeV] = 0.73\cdot\sqrt[3]{\frac{\rho}{\rho_0}};~~(sph)~~~~~~  \nonumber
\end{eqnarray}
where D is the number of $d^*(2380)$, $\rho$ is the average condensate density and $\rho_0$ is nuclear matter density. The coefficients $a_{v}'$ and $a_{c}'$ determine the relative strengths of the attractive volume and the repulsive Coulomb terms (see Appendix A1).
The magnitude of these coefficients for the case of a BEC will be dictated by the average relative separations of the constituents and the strength of the strong interactions between them. The isoscalar nature of the $d^{*}(2380)$ interaction gives expectation of a force somewhat weaker than the isovector N-N interaction. This is already established for the analogous (isoscalar)  $\Lambda$-$\Lambda$ hyperon interaction, where the scattering length is an order of magnitude weaker than the corresponding N-N.  Recent calculations of the range of the $d^*-d^*$ interaction~\cite{ale2019} ~\footnote{Due to the isoscalar nature of the $d^{*}(2380)$, isovector meson exchanges (e.g. $\pi$, $\rho$) are excluded and higher mass meson exchanges (e.g. $\eta$, $\eta^{’}$ mesons) are known to have weak coupling. Therefore, estimates based on a modified NN potential, accounting for the higher $d^{*}(2380)$ mass and assuming a dominance from the remaining $\sigma$-meson (scalar-isoscalar two-pion exchange) and $\omega$-meson exchange in the interaction and a meson-$d^{*}(2380)$ coupling a factor 2 larger than meson-$N$ coupling to account for the 6-quark rather than 3-quark systems. This results in a force showing suppression of longer range interactions and strong short range attraction with even stronger shorter range repulsion. The detailed calculations will be published in a forthcoming paper~\cite{ale2019}.} also indicate an interaction of smaller magnitude and shorter range.    
  We take $a_v'$ over all values from 0 up to the NN limit of 30 MeV to to explore the sensitivity of the condensate properties to the detailed nature of the $d^{*}(2380)$ interaction. We note that assuming an interaction strength scaled from the $\Lambda\Lambda$-case, and consistent with recent $d^*d^*$ interaction calculations, would give $a_v'$ of order $\sim1$~MeV.  As shown in equations 2, the size of $a_c'$ has a weak dependence on matter density. The range of $a_c'$ covers densities up to $\rho=10\rho_0$, reflecting the expected shorter range of the interaction compared to the NN case. However, as will be shown the results are rather insensitive to $a_c'$ over this wide range of densities. Experiments on $d^*(2380)$ production in the nuclear medium, will help to determine the strength of the $d^*(2380)$ interactions. Such experiments are planned to be done at CB@MAMI, JLab and could be performed at several other facilities, like FAIR@GSI. An exploratory study was also performed at WASA~\cite{APr2012}. 


Fig 1 shows a plot of the predicted binding energy of the $d^{*}(2380)$-BEC as a function of $d^{*}(2380)$ multiplicity and $a_v'$. The results provide an indicative parameter space for a bound $d^{*}(2380)$-BEC, defining regions where the energy required to form the $d^*(2380)$ from nucleons is exceeded by the binding energy available in forming the $d^*(2380)$-BEC. For the range of $a_{v}'$ calculated (corresponding to a range of potential attractive/repulsive $d^*$-forces) the {\it minimum} multiplicity threshold for a stable $d^{*}(2380)$-condensate ranges from the order of hundreds (mass in the TeV range) to the order of millions. For the previously motivated value of $a_v'\sim 1$~MeV a multiplicty threshold of around 1000 $d^{*}(2380)$ is predicted for a stable condensate. Ongoing experimental programmes aiming to better determine the $d^{*}(2380)$ interaction in nuclear matter may allow tighter constraints on this multiplicity threshold.  The small relative contribution of the Coulomb term compared to the volume term means the conclusions are rather insensitive to the adopted value of $a_{c}^{'}$ or to the assumption of a linear-chain configuration. The latter point is illustrated in Fig. 1 where the bound threshold for a spherical condensate is also presented for 3 values of condensate density, $\rho=3,5$ and $10\rho_0$. Significant deviations from the  linear-chain configuration calculation,are only visible for regions where $a_v'$ is small.

\begin{figure}[!h]
\begin{center}
  \includegraphics[width=0.7\textwidth,angle=0]{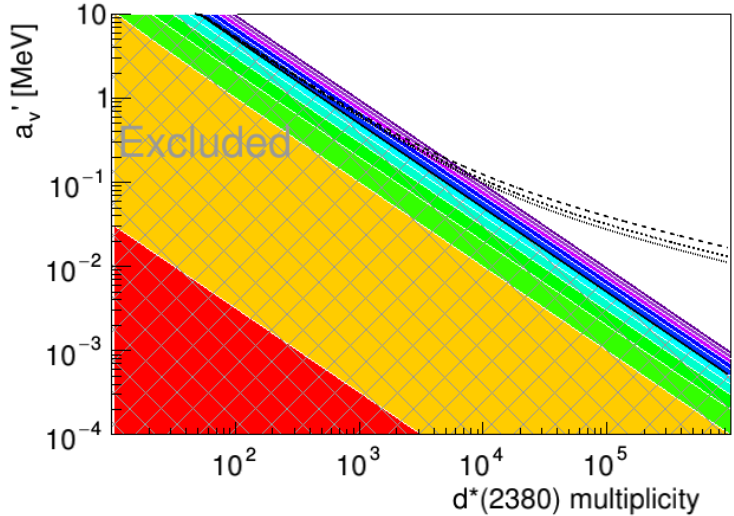}
\end{center}
\caption{The $d^{*}(2380)$ binding energy (z-axis) presented as a function of the multiplicity and the magnitude of  $a_v'$ (see text). The binding energy contours are shown from 500~MeV unbound (red) to 500 MeV bound (dark violet) in intervals of 100~MeV. Unbound regions are identified by the hatching. Solid line correspond to the bound threshold for linear condensates, assuming $\rho=5rho_0$, while dotted, short-dashed and long-dashed lines represent spherical condensate bound thresholds with  $a_c'$ corresponding to $\rho=3,5$ and $10\rho_0$ respectively.  Note the double-logarithmic scale.}
\label{avgr}
\end{figure}

In order to give further validity to the results, we can check for consistency with other theoretical ansatz employed in BEC studies. The $d^{*}(2380)$-BEC proposed here is in somewhat analogous to the (few-particle) $\alpha$-BEC attributed\cite{ACN0,ACN} to the Hoyle states in $^{12}C$ and $^8Be$. Predictions in the Gross-Pitaevskii formalism (GPF) for a dilute isoscalar $N\alpha$-boson BEC give predicted scalings of binding energy with multiplicity consistent with those in equation 2 (the GPF formalism gives total kinetic energy $T\propto (N-1)$ and two-body potential energy $V\propto N(N-1)/2$).

Estimates of upper limits of $d^{*}(2380)$-BEC stability can be obtained from recent generic formulations for the properties of BEC formed from fundamental particles, which additionally include gravitational binding. For a recent overview on self-interacting BEC dark matter see {\it e.g.} Ref.\cite{BEC}. The maximum BEC mass for an attractive condensate is $M_{max}=\frac{\hbar}{\sqrt{|a|Gm}}$, \cite{BEC}, where {\it a} is the scattering length, {\it m} is the particle mass and {\it G} is the gravitational constant. Using $M_{d^*} = 2.38$ GeV and $|a|\sim$1~fm gives $M_{max}\sim 10^{17}$ GeV, corresponding to a $d^{*}(2380)$-BEC having size and mass limits of order Angstroms and grams respectively, characteristic DM scales well studied for the analogous "WIMPzilla" scenarios\cite{WIMPzilla}.

\section{\label{sec:astro}Quantification of primordial production}

 For $d^{*}(2380)$-BECs to be a plausible new DM candidate then a site with the potential to produce sufficient quantities should be established. An estimate of $d^{*}(2380)$-BEC primordial production in the early universe can be obtained from thermodynamic considerations and in our estimates we follow the methodology outlined in Ref.~\cite{Farrar}. The most significant primordial production would be expected during the transition from quark-gluon plasma to hadrons and this was estimated as a function of decoupling temperature and $d^{*}(2380)$-BEC binding energy per baryon (or equivalently its mass) (see Appendix A2). The predicted primordial abundance of $d^{*}(2380)$-BEC as a ratio to conventional baryonic matter is presented as a function of these quantities in Fig. 2. It is clear that for a wide range of (feasible) $d^{*}(2380)$-BEC binding energies the predicted primordial abundance greatly exceeds that of standard baryonic matter. At the decoupling temperature of the quark-gluon plasma (QGP) to hadrons ($\sim$155~MeV) very significant primordial production of $d^{*}(2380)$-BEC is evident for a wide range of binding energies. Also displayed on the figure is the experimentally determined  constraint on the dark matter (DM) fraction in the universe; $\Omega_{DM}/\Omega_{baryon}=5.3\pm 0.1$. Assuming dominant freeze out close to the temperature of the QGP phase transition ($\sim$155 MeV), and taking the limit where {\em all} observed DM arises from $d^{*}(2380)$-BECs, would necessitate a binding energy of order $\sim$200~MeV (see Fig.~1 for the corresponding constraints on condensate multiplicities). This could be reduced even further if condensate formation would also be possible at lower decoupling temperatures in the hadronic phase, for example through $pn\leftrightarrow d^*(2380)$ processes.  

The concept of a new, light-quark QCD inspired DM candidate produced near the QGP phase transition is also  interesting due to the recently proposed separation of quarks and antiquarks at freezeout\cite{CP1}. The signatures of such CP violating effects are searched for in a number of ongoing heavy-ion beam experimental programmes. If this is confirmed, the possibility for a preferential formation of antimatter $\bar{d}^*(2380)$-BECs is worthy of further study due to the possibility to "hide" primordially produced antimatter in the form of a BEC.

\begin{figure}
\begin{center}
\includegraphics[width=0.7\textwidth,angle=0]{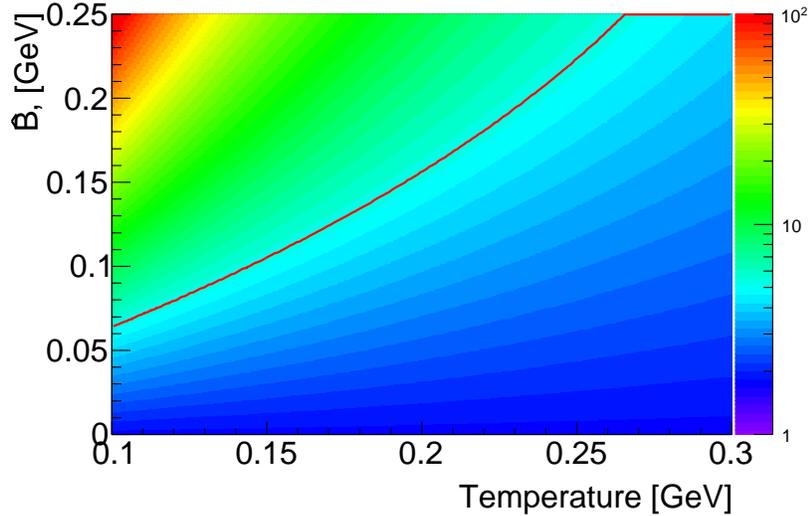}
\end{center}
\caption{The primordial production of $d^*(2380)$-BEC (expressed as a ratio to baryon matter) calculated as a function of binding energy per baryon, $\hat{B}$(see App. A2 for details) and decoupling temperature. The red line shows the loci corresponding to the current experimental determination of the dark matter to matter ratio\cite{CMB}.}
\label{abund}
\end{figure}

\section{\label{sec:astro1}Potential astronomical signatures}

The theoretical estimates for $d^*(2380)$-BEC formation indicate it could be a plausible new light-quark based DM candidate. As creating the necessary $d^{*}(2380)$ multiplicities and densities for BEC formation (see sec. 2) is out of the reach of currently conceivable experimental facilities, then confirmation of the existence of  $d^*(2380)$-BECs relies on astrophysical signatures. In this section we employ our current knowledge of the $d^{*}(2380)$'s properties to identify possible features that could be incorporated in future DM searches.

As discussed in section 2 our results indicate $d^*(2380)$-BEC have a formation threshold of around $\sim$1000 constituents if the $d^*-d^*$ interaction scales simply from the nucleon-nucleon case,  or a threshold of many millions if the $d^*-d^*$ interaction is weaker. The upper $d^*(2380)$-BEC multiplicity limits (for the assumed attractive $d^*-d^*$ interaction) correspond to masses at the  gram scale. Such $d^*(2380)$-BEC "nuclei" would posses a very large positive charge. As discussed in section 2 the expected absence of such a surface effect in the $d^*(2380)$-BEC favours "chain-state" structures, in which the Coulomb repulsion between constituents is minimised. Such behaviour is already observed for the analogous $\alpha$-BEC "chain states" in light nuclei\cite{alcon}. Neutral atomic systems may form around the $d^*(2380)$-BEC, but the equivalent atomic number would be well in excess of standard atoms. Calculations of, and subsequent searches for, atomic line series from such high-Z, extended structures may provide one route to the observation of $d^*(2380)$-BECs, if transitions in the $d^*(2380)$-BEC atomic series are distinguishable from, and inconsistent with, known atoms. With recent advances in X-ray telescopes anomalous  transitions are already being identified~\cite{XRay}.

However, if larger mass and therefore highly-charged $d^*(2380)$-BEC form, the bound electrons (and muons if the Fermi momentum is sufficiently high) may have such confined radii that the structure tends towards mixed hadronic and leptonic matter.  Such heavy "atoms" would have a net zero charge while being gravitationally weak. Therefore observation through standard astronomical methods based on atomic electromagnetic signatures may be challenging or impossible.

\begin{figure}
\begin{center}
\includegraphics[width=0.7\textwidth,angle=0]{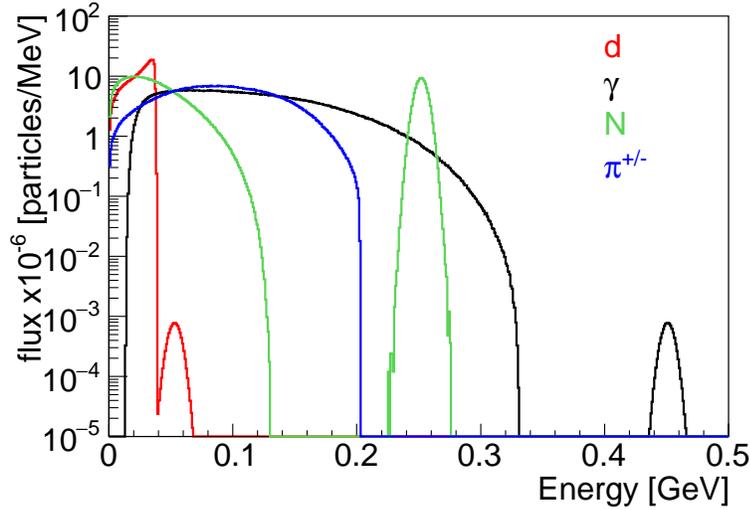}
\end{center}
\caption{The $d^{*}(2380)$-BEC decay spectrum, presented as a flux per $d^*(2380)$, per MeV of kinetic energy. The flux is separated into the contributions from gamma(black), nucleons(green), deuterons(red) and charged pions(blue).}
\label{DDec}
\end{figure}

The direct observation of decays products from a $d^*(2380)$-BEC are likely to be the clearest experimental signature. Condensate breaking would require an energy transfer to the condensate (e.g via cosmic ray interactions) greater than the gap between the condensed and non-condensed states. Fig. 3 shows the expected yield of reaction products from a complete condensate collapse of a $d^*(2380)$-BEC at rest, calculated using the established "free-space" decay branches of the $d^*(2380)$. A very high multiplicity, localised and essentially instantaneous emission of $\gamma$ radiation along with an intense flux of deuterons, nucleons and charged pions would be expected from a condensate collapse\cite{decays}. Such characteristic emission spectra could be searched for in astronomical observations.  

The existence of $d^*(2380)$-BEC decays in the earth's atmosphere or close to its surface would produce energies comparable with cosmic-ray events, but {\em without} directionality. As cosmic-rays should not be able to pass through the earth based on our current understanding of standard model physics, then upward going "cosmic-ray" events may provide a potential signal. The experimentally observed (and currently unexplained) upward going cosmic-ray events\cite{darkRay} could also therefore be explored to put astronomical constraints on $d^*(2380)$-BECs. 
 
 An enhancement of any $d^*(2380)$-BEC decay signal could be achieved if signatures of the parent $\pi^{0}$ mesons, which dominantly produce the $\gamma$ flux, could be identified. Such measurements are already obtained for $\pi^{0}$ produced from the moon's surface (e.g. with the Fermi-LAT telescope\cite{Moon1}). However, better directional information and proximity to the surface to localise the source would be of benefit, such as may be achieved in the Lunar Prospector mission\cite{Moon2}. 

\section{Conclusion}

In this paper we present a first investigation into the feasibility of the formation of stable Bose-Einsten condensates of bosonic, isoscalar, hexaquark matter formed from the recently discovered $d^*(2380)$. The $d^*(2380)$ condensates are predicted to be stable for a multiplicity threshold ranging from 1000's to millions dependent on the detailed nature of the $d^{*}(2380)$'s self-interaction. Very significant primordial production at the quark-gluon plasma to hadronic phase transition appears feasible dependent on the condensate properties, suggesting a new potential candidate for dark matter. Some possible astronomical signatures and search scenarios for establishing the existence of such matter are presented. The ongoing and planned X-ray and gamma ray astronomy facilities could potentially set limits on the existence of $d^*(2380)$ condensates. 
\section{Acknowledgement}
This work has been supported by the U.K. STFC ST/L00478X/1 and ST/P004008/1 grants.

\section{Appendix A1}
The derivations of the form of the Coulomb terms and their density dependence for the case of spherical and linear condensates are outlined below.
\subsection*{Spherical case}

The electrostatic energy $E$ of a spherical nucleus can be approximated by the energy of a homogeneously charged sphere enclosing a charge, Q:

\begin{eqnarray}
E=\frac{3}{5}\frac{1}{4\pi \epsilon_0}\frac{Q^2}{R}\approx \frac{3}{5}\frac{1}{4\pi \epsilon_0}\frac{Z(Z-1)}{r_0 A^{1/3}} 
\end{eqnarray}
where $Z$ is the nuclear charge and $r_0$ the parameter defining the average separation of nucleons in the nucleus (or equivalently hexaquarks in a condensate). The $r_0$ parameter is extracted from the observed scaling of nuclear radii with mass number  $R=r_0A^{1/3}$ (or for condensate $R=r_0D^{1/3}$). From comparison with the form of the Coulomb term in equation 1, $a_c$ is given by:

\begin{eqnarray}
a_c=\frac{3}{5}\frac{1}{4\pi \epsilon_0}\frac{e^2}{r_0}=\frac{3}{5}\frac{\hbar c \alpha}{r_0} \\
a_c [MeV]=\frac{0.86}{r_0 [fm]} \\
\end{eqnarray}
where $\hbar$ is the reduced Planck's constant, $e$ the electron charge, $\alpha$($\sim 1/137$) the fine structure constant and $c$ the speed of light in vacuum. The standard nuclear value ($r_0=1.25$~fm) gives $a_c=0.7 MeV$. The fomula for $a_c$ can also be expressed as a function of the dimensionless parameter $\rho$/$\rho_0$:
\begin{eqnarray}
\rho=\frac{m_N A}{4/3\pi r_0^{3} A} \to r_0=\frac{1.18 [fm]}{\sqrt[3]{\rho [\rho_0]}} \\
{\bf a_c' [MeV] = a_c [MeV] = 0.73 \cdot \sqrt[3]{\rho/\rho_0}}
\end{eqnarray}
This way the formula for the $a_c$ calculations can be generalised as a function of the matter density, including for the case of condensate matter. 
Note that for the spherical case $a_c'$ and $a_c$ (see text) are the same. Equation 8 illustrates the weak dependence of the $a_c$ term with matter density. 

\subsection*{Linear chain case}

The Coulomb energy for a homogeneously charged rod of enclosed charge (Q), having length l is given by electrostatic theory as:

\begin{eqnarray}
E=\frac{1}{4\pi \epsilon_0}\frac{Q^2}{l} \\
a_c'=\frac{\hbar c\alpha}{l} \\
a_c' [MeV]=\frac{1.43}{l [fm]} 
\end{eqnarray}

For the linear case the volume is defined by the width and the length of the cylindrical volume. Therefore an additional parameter is needed, the radius of the $d^*(2380)$.  Here we adopt a value of $r_{d^*}=0.5fm$ in accordance with Ref.14. However, $a_c'$ has a rather weak dependence on the adopted value. 
\begin{eqnarray}
\rho=\frac{m_{d^*} D}{\pi r_{d^*}^{2} l D} \to l=\frac{5.6 [fm]}{r_{d^*}^2[fm^2]\cdot \rho [\rho_0]} \\
r_{d^*}=0.5 [fm] \to l=\frac{22.4 [fm]}{\rho [\rho_0]} \\
{\bf a_c' [MeV] = 0.064 \cdot \rho/\rho_0}  
\end{eqnarray}

For the linear condensate the binding energy contribution per $d^*$ from the Coulomb term is therefore constant i.e. independent of the number of participating $d^*$'s. Note that this differs from the volume term, which is proportional to the number of participants.

\section{Appendix A2}
Boltzman's law give the number density of particles with mass $m$, degrees of freedom ($g$) as:
\begin{eqnarray}
n=g\left(\frac{mT}{2\pi}\right)^{3/2}e^{\frac{\mu-m}{T}}
\end{eqnarray}
where $T$ is the temperature and $\mu$ is the chemical potential.
For a non-strange medium at high temperature ($m_d-m_u<<T$), the number of $u$ and $d$ quarks approaches equality. For this situation it is possible to  compare the number density of the proposed "hexaquark matter" with "proton-neutron" matter.

The $d^*$ has spin $S=3$, giving $2S+1$ degrees of freedom and $g_{d^*}=7$.
For the nucleonic case the two nucleons can be either in spin $S=1$ or $S=0$ states. Summing these possibilities  gives $g_{pn}=4$. Taking $\hat{m_{d^*}}$ as the mass of the bound $d^*$ in condensate gives:
\begin{eqnarray}
\frac{n_{d^*}}{n_{pn}}=\frac{7}{4}\left(\frac{\hat{m_{d^*}}}{2m_N}\right)^{3/2}e^{\frac{2M_N-\hat{m_{d^*}}}{T}}
\end{eqnarray}

Introducing the binding energy of the $d^*$ in condensate per baryon as $\hat{B}=\frac{2M_N-\hat{m_{d^*}}}{2}$ we can rewrite equation 16 with a matter density parameter $\Omega$ as:
\begin{eqnarray}
\frac{\Omega_{DM}}{\Omega_{matter}}=\frac{7}{4}\left(\frac{2m_N-2\hat{B}}{2m_N}\right)^{5/2}e^{\frac{2\hat{B}}{T}}=\frac{7}{4}\left(1-\frac{\hat{B}}{m_N}\right)^{5/2}e^{\frac{2\hat{B}}{T}}
\end{eqnarray}

\end{document}